\documentclass[aps,prl,twocolumn,showpacs]{revtex4}
\usepackage{epsfig}

\newcommand{\myth}{{\rm tanh}}
\newcommand{\cth}{{\rm coth}}

\newcommand{\myRe}{{\rm Re}}

\newcommand{\gton}{\mathrel{\lower.9ex \hbox{$\stackrel{\displaystyle 
>}{\sim}$}}} 
\newcommand{\lton}{\mathrel{\lower.9ex \hbox{$\stackrel{\displaystyle 
<}{\sim}$}}}  

\newcommand{\vB}{{\bf B}}
\newcommand{\vC}{{\bf C}}
\newcommand{\vE}{{\bf E}}

\newcommand{\vp}{{\bf p}}
\renewcommand{\vr}{{\bf r}}

\begin{document}

\title{Momentum anisotropy in nuclear collisions from quantum mechanics}

\author{Denes Molnar}
\author{Chris H.~Greene}
\author{Fuqiang Wang}
\affiliation{Department of Physics and Astronomy, 
Purdue University, West Lafayette, IN 47907}

\date{\today}

\begin{abstract}
We point out that the intrinsic relationship between space and momentum in 
quantum physics through the uncertainty principle has potential implications 
for momentum anisotropy in heavy-ion collisions. Using a harmonic oscillator 
potential we calculate the elliptic anisotropy and find it to be sizeable
 compared to elliptic flow measurements in nuclear collisions. Our results 
question the validity of the completely hydrodynamic interpretation of 
anisotropic flow data, and highlight the importance of including quantum 
physics in hydrodynamic calculations which has largely been neglected so far.
\end{abstract}


\maketitle

{\em Introduction.} 
One of the most spectacular observations in high-energy nuclear collisions is 
the so-called elliptic flow~\cite{v2exp}. This ``left-right'' versus 
``up-down'' momentum asymmetry of particle production in the plane transverse
to the colliding beams, commonly characterized by a Fourier coefficient
$v_2 = \langle \cos 2\phi \rangle$, is understood as the collective
dynamical response of hot and dense plasma of quarks and gluon created 
in a {\em spatially} asymmetric overlap region
of target and projectile in the collision. In the 
hydrodynamic picture~\cite{hydro}, 
elliptic flow is generated by spatially asymmetric
pressure gradients in the system, resulting in nonzero final momentum 
anisotropy ($v_2$) even if the initial momentum distribution is
isotropic azimuthally.

Recently, a similar anisotropy has been seen in proton-nucleus~\cite{pA_v2} 
and even
high-multiplicity proton-proton collisions~\cite{pp_v2} at the Large Hadron
Collider. Hydrodynamic calculations~\cite{pp_pA_hydro}
of these collisions appear to qualitatively
and semi-quantitatively reproduce the observations.
It is astonishing, however, that a macroscopic theory like hydrodynamics
applies so well at such short length and time scales of ${\cal O}(1 {\rm\ fm})$. 
One might expect quantum mechanics to also play a role.
The intrinsic quantum connection between momentum and coordinate spaces
through the Heisenberg uncertainty relation
implies that spatially anisotropic
systems are in general anisotropic in momentum as 
well. It has been stressed in atomic Bose-Einstein condensation studies,
for instance, that this is responsible for causing momentum distributions 
to have the opposite anisotropy compared to the trapping potential anisotropy
(see, e.g., Ref.~\cite{Anderson}).

The following estimates this intrinsic momentum asymmetry
for thermally equilibrated particles trapped in an anisotropic potential. 
We simplify the analysis by assuming nonrelativistic motion in
either an exactly solvable harmonic oscillator or else a 
square well potential.
In addition, we solve the case of ultrarelativistic (massless particle) motion, numerically, 
in a harmonic oscillator potential. 
The results are inevitably qualitative but still provide insight into the
possible origin and magnitude of 
momentum anisotropy in hadron and nuclear collisions.

{\em Momentum anisotropy from quantum mechanics.}
Consider a thermally equilibrated system of independent particles in an 
anisotropic potential
described by a Hamiltonian
$H = \sum\limits_{n=1}^{N} {H_1}(\vp_n,\vr_n)$,
with separable
$H_1(\vp,\vr) = K(\vp) + V(\vr)$ for simplicity.
Assume that the system 
begins in thermal equilibrium with particles confined
in two dimensions due to the potential, with the
$z$ direction ignored~\cite{note_z}.
The trap is then suddenly removed ($V \to 0$).

Classically, the momentum distribution~\cite{note_unit},
$
\frac{dN}{d\vp}= N \frac{\int d\vr\, e^{-H_1(\vp,\vr) / T}}
                        {\int d\vr\, d\vp\, e^{-H_1(\vp,\vr) / T} }
       = N \frac{ e^{-K(\vp) / T}} {\int d\vp\, e^{-K(\vp) / T} }\,,
$
is independent of the potential $V$, and so it
is isotropic at all temperatures 
as long as $K$ is isotropic. 
Quantum mechanically, however, the same system exhibits momentum anisotropy
in general {\em even if the kinetic energy $K$ is isotropic}. 
This is apparent in the $T \to 0$ single-particle limit
($N=1$) where the uncertainty relation implies that
$\Delta p_x \sim 1/\Delta x$, 
$\Delta p_y \sim 1/\Delta y$ for the ground state.

{\em Single-particle system.} 
For one particle ($N=1$) at temperature $T$
the spatial density is given by the canonical average
\begin{equation}
\rho(\vr) \equiv \frac{dN}{d\vr} = \frac{1}{Z}
         \sum_j |\psi_j(\vr)|^2\, e^{-E_j/T} 
\label{rho2D}
\end{equation}
with $Z \equiv \sum_{j} e^{-E_j/T}$. The sums
are over the complete orthonormal set of eigenstates of $H_1$,
and $E_j$ and $\psi_j(\vr) = \langle \vr | j\rangle$ with $\vr=(r_x,r_y)$ 
are the energy and coordinate-space
wave function for eigenstate $|j\rangle$.
The distribution of momenta $\vp = (p_x,p_y)$ 
is given by the analogous sum 
\begin{equation}
f(\vp) \equiv \frac{dN}{d\vp}
       = \frac{1}{Z} \sum_j |\psi_j(\vp)|^2 \, e^{-E_j/T}  \ ,
\label{f2D}
\end{equation}
where $\psi_j(\vp) = \frac{1}{2\pi} \int d\vr \, e^{-i\vp \vr} \psi_j(\vr)$
is the Fourier transform of $\psi_j(\vr)$.
It is straightforward to show that $f(\vp)$ is unaffected by
the free evolution $\psi \to \exp(-i \hat K t)\;\!\psi$ 
of the ensemble of thermal states after the trap is removed.

Next consider the nonrelativistic $K(\vp) = (p_x^2 + p_y^2) / 2M$ with
a harmonic oscillator potential 
$V(\vr) = M(\omega_x^2 r_x^2 + \omega_y^2 r_y^2) / 2$,
where $M$ is the particle mass.
The momentum anisotropy is characterized by the parameter
$\bar{v}_2 \equiv \langle\cos(2\phi)\rangle=\langle p_x^2-p_y^2\rangle / \langle p_x^2 + p_y^2\rangle$,
and the initial spatial eccentricity by
$\varepsilon \equiv \langle r_y^2 - r_x^2 \rangle/\langle r_x^2 + r_y^2 \rangle$.
At temperature $T$ the averages that appear in $\bar v_2$ and $\varepsilon$
are readily~\cite{note_expectation} calculated $(i=x,y)$:
\begin{equation}
    \langle p^2_i\rangle
    = \frac{M\omega_i}{2}\, \cth\frac{\omega_i}{2T} 
 \ , \quad
\langle r_i^2 \rangle = \frac{1}{2M\omega_i}\, \cth \frac{\omega_i}{2T}\,.
\label{HO_sizes}
\end{equation}

\begin{figure}
\begin{center}
\hspace*{0mm}\includegraphics[height=40mm]{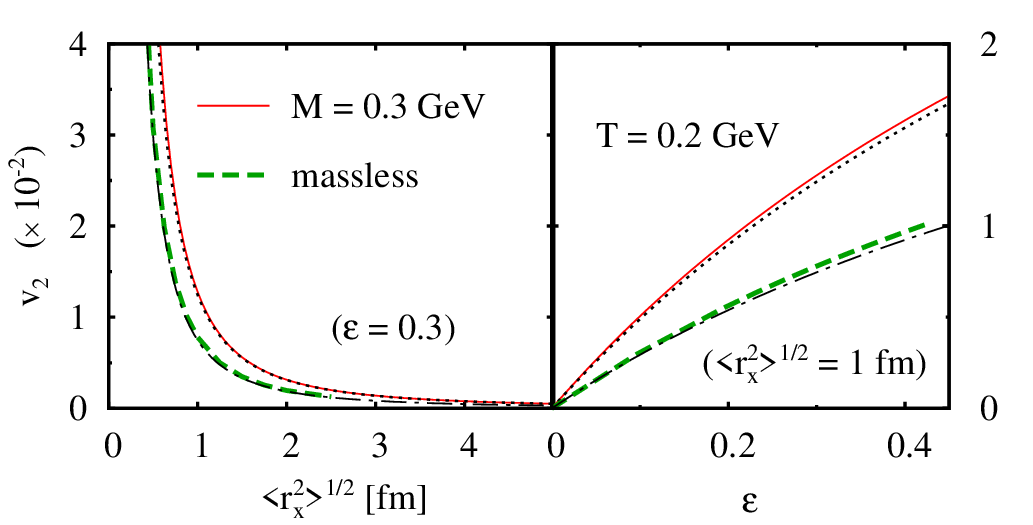}
\end{center}
\caption{Average elliptic anisotropy $\bar v_2$ versus mean system size 
$\langle r_x^2\rangle^{1/2}$ for spatial eccentricity $\varepsilon=0.3$ 
(left), 
and $\bar v_2$ versus $\varepsilon$ 
for $\langle r_x^2\rangle^{1/2}=1$~fm (right). 
Both plots are for $T=0.2$~GeV with $M=0.3$~GeV (solid lines)
or zero (dashed). Approximate results from Eq.~(\ref{HOv2_approx}) (dotted) and Eq.~(\ref{HOv2_approx}) with 
$M=2.5T$ (dashed-dotted) are also shown.
}
\label{fig:avv2_vs_L_eps}
\end{figure}

Figure~\ref{fig:avv2_vs_L_eps} 
illustrates (solid curves) the spatial trap size and eccentricity dependence, general features of the average elliptic anisotropy,
for particle mass $M=0.3$~GeV and temperature $T=0.2$ GeV.
This mass value can be thought of as 
a constituent quark, or a particle with mass $M \sim T$.
Here $\bar v_2$ drops monotonically with system size,
but increases monotonically with eccentricity. 
For a system of $\langle r_x^2\rangle^{1/2}\sim 1$~fm size, $\bar v_2 \sim {\cal O}(10^{-2})$.
Dotted lines correspond to the remarkably accurate approximation 
in Eq.~(\ref{HOv2_approx}).

It is instructive to see how the classical limit is approached at
high temperature.
The relevant parameters are $\omega_i/T$, namely the
ratios of the energy level spacing to the typical energy. 
For $\omega_i \lton T$ Taylor expansion of
Eq.~(\ref{HO_sizes}) directly relates the average anisotropy 
to the spatial eccentricity $\varepsilon$:
\begin{equation}
\bar v_2 \approx \frac{\hbar^2 }{12 k_BTM \langle r_x^2 \rangle}\cdot\frac{\varepsilon}{1+\varepsilon}
\label{HOv2_approx}
\end{equation}
using traditional units for clarity.
The same $\Delta p_x  \Delta x / \hbar$ combination
appears here as in the uncertainty relation,
since near the classical limit
$\sqrt{\langle p_x^2\rangle} \approx \sqrt{k_B T M} \equiv p_{\rm th}$
is just the typical thermal momentum.
The average elliptic anisotropy is 
roughly one-tenth of the eccentricity
when $p_{\rm th}^2\langle r_x^2\rangle = \hbar^2$.

A crosscheck of these features using an infinite square-well potential
shows that the onset of clasical behavior is 
controlled by the same scale $p_{\rm th}^2 \langle r_x^2\rangle$,
but unlike the harmonic oscillator results the square root 
of the scale now appears. For example, near the classical
limit
\begin{equation}
\bar v_2 \approx \sqrt{\frac{\pi}{96 TM \langle r_x^2 \rangle}}
\left(1 - \sqrt{\frac{1-\varepsilon}{1+\varepsilon}}\right)
         \approx \frac{0.18}{\sqrt{TM \langle r_x^2 \rangle}}
          \frac{\varepsilon}{1+\varepsilon / 2} \,.
\end{equation}
The reason behind the slower approach is that for the infinite square well 
$TM\langle r_x^2 \rangle$
corresponds to the ratio of temperature to gap between the lowest two 
energy states of the system; unlike the harmonic oscillator
this is not indicative of the level spacing compared to $T$ 
(since levels grow quadratically).

Figure~\ref{fig:avv2_vs_L_eps} also shows (dashed curves) the quantum anisotropy for a massless 
(ultrarelativistic) particle trapped in the anisotropic Gaussian potential,
for which $K = (p_x^2 + p_y^2)^{1/2}$.
The results were obtained by
numerically solving the eigenvalue problem for the Hamiltonian
in a large computational basis ($\sim 10^5$ states), 
and using Eq.~(\ref{f2D}) for $f(\vp)$
(details of the computation will be presented elsewhere). 
On general grounds, $\bar v_2$ for massless particles
is expected to be a decreasing function of the product of temperature and system size,
and an increasing function of eccentricity. Our results not only match these expectations
but show very similar system size and eccentricity dependence to the nonrelativistic $\bar v_2$, 
and a comparable magnitude as well.
Interestingly, the nonrelativistic formula
Eq.~(\ref{HOv2_approx}) captures
the massless particle $\bar v_2$ very well if one sets $M = 2.5T$ (dashed-dotted lines).

A more differential measure of momentum anisotropy is
provided by the Fourier coefficients
$v_n(p_T)\equiv\langle\cos n\phi\rangle_{p_T}=
\int_0^{2\pi} d\phi\,\cos n\phi\, f(p_T,\phi)
\left/\int_0^{2\pi} d\phi\, f(p_T,\phi)\right.
$,
where $p_T = \sqrt{p_x^2+p_y^2}$ and $\phi= \arctan(p_y/p_x)$ 
are the magnitude and azimuth of the momentum vector. 
In particular, $v_2(p_T)$ describes the momentum dependence of ``elliptic''
anisotropy. {\em For nonrelativistic particles}, the calculation of $v_n(p_T)$ is straightforward,
so we only highlight a few key points.
Eigenstates are labelled by a pair
of nonnegative integers, $|j\rangle \equiv |n m\rangle$,
the wave function factorizes
$\psi_{nm}(\vr) = \psi_n(r_x) \psi_m(r_y)$, and energy is additive
$E_{nm} = E_n + E_m = (n+\frac{1}{2})\omega_x + (m+\frac{1}{2})\omega_y$.  
Therefore, 
the canonical partition sum, 
wave function Fourier transform,
momentum distribution, and
particle density all factorize, also for the square well, i.e., $Z = Z_x Z_y$, 
$|\psi_{nm}(\vp)|^2 = |\psi_n(p_x)|^2 |\psi_m(p_y)|^2$,
$f(\vp) = g_x(p_x) g_y(p_y)$, and $\rho(\vr) = \rho_x(r_x) \rho_y(r_y)$.
In Eq.~(\ref{f2D}) the Fourier transform 
involves the same Hermite polynomials as those in the 
coordinate-space
eigenstates~\cite{note_rep}, 
while the canonical sums of $|\psi_n(x)|^2$ and $|\psi_n(p)|^2$
are doable using Mehler's formula~\cite{Mehler}.

The density and momentum distributions both 
turn out to be (asymmetric) Gaussians
\begin{equation}
\rho(\vr) \propto \exp\!\left(-\sum_{i} \frac{r_i^2}{2\langle r_i^2\rangle}\right) \,,
\quad
f(\vp)    \propto \exp\!\left(-\sum_{i} \frac{p_i^2}{2\langle p_i^2\rangle}\right) \,.
\label{f_rho_HO}
\end{equation}
An intuitive reason for why Gaussians appear is
that for $\omega_x = \omega_y$ there is azimuthal
symmetry, and a product form such as $f(\vp) = g(p_x) g(p_y)$ is
azimuthally symmetric only for Gaussian $g$~\cite{note_gaus}.

In the $T\to \infty$ limit momentum anisotropy vanishes
because both $\langle p_x^2\rangle$ and $\langle p_y^2\rangle$ approach $TM$ 
and we recover the classical
Maxwell-Boltzmann distribution.
At any finite $T$, however,
the tail of the momentum distribution is always
anisotropic. Deviations from classical Maxwell-Boltzmann
are a factor of two or more when either momentum component 
exceeds approximately $\sqrt{24\, \ln 2}\, \, p_{\rm th} T/ \omega \approx 
4 \sqrt{TML^2}\, p_{\rm th}$,
where $L^2$ and $\omega$ are the mean-squared system size and trap strength
in the 
corresponding direction.
The system is classical over a wider range of $p/p_{\rm th}$, 
the heavier the particle, 
the larger the temperature, 
or the larger the system size.

Differential harmonic coefficients readily follow using Eq.~(\ref{f_rho_HO}):
\begin{equation}
v_{2n}(p_T) 
   = h_n\!\left(\frac{p_T^2}{2MT} (S_y - S_x)\right)
\ , \quad 
     S_i \equiv \frac{T}{\omega_i} \myth\frac{\omega_i}{2T}\ ,
\label{vn_HO} 
\end{equation}
where $h_n(x) \equiv I_n(x) / I_0(x)$ is a ratio of 
modified Bessel functions of the first kind.
Even harmonics are all monotonic in $p_T$, start out at low $p_T$ as 
$v_{2n} \propto p_T^{2n}$, and approach unity at infinite momentum. 
A useful approximation for low $p_T$ and $\omega_i \lton T$ is
$v_2(p_T) \approx {\bar v}_2\,  p_T^2/4MT$, i.e.,
$\bar v_2 \approx v_2(2p_{\rm th})$ (note the factor $4MT$ and not $2MT$ here).
Also, $v_{2n}(p_T)\approx [v_2(p_T)]^n/n!$ in the same limit.
Odd harmonics vanish by symmetry for our even anisotropic trap,
but those can be produced with a less symmetrical anisotropic trap.

\begin{figure}
\begin{center}
\hspace*{-2mm}\includegraphics[height=40mm]{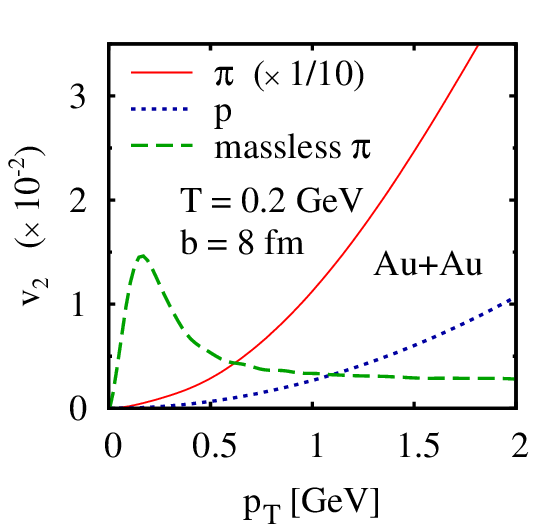}
\hspace*{2mm}\includegraphics[height=40mm]{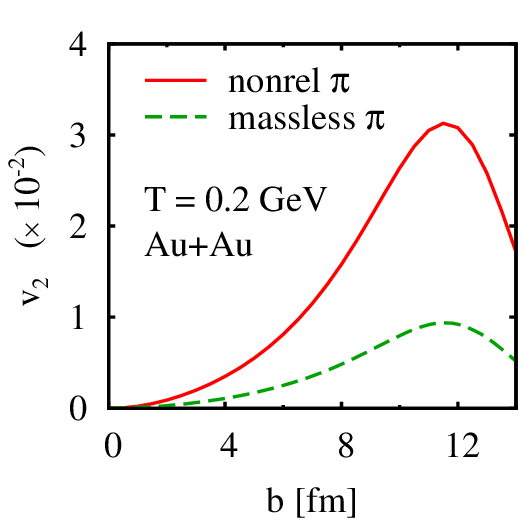}
\end{center}
\vspace*{-0.5cm}
\caption{{\em Left:} Elliptic anisotropy versus transverse momentum for a 
system of pions (solid, scaled by 1/10) and protons (dotted). 
{\em Right:}
Average elliptic anisotropy versus impact parameter ($b$) for Au+Au collisions.
The result for massless particles is also shown in both panels (dashed).
The binary collision transverse profile is used to set mean-squared dimensions of the initial collision zone of Au+Au system. For the left panel $\langle r_x^2\rangle^{1/2}=1.5$~fm and $\langle r_y^2\rangle^{1/2}=2.2$~fm corresponding to $b=8$~fm. For both panels $T=0.2$~GeV is used.}
\label{fig:rhicv2}
\end{figure}

Numerical results for $v_2(p_T)$ in the massless limit
are discussed in the next section (cf. Fig.~\ref{fig:rhicv2}).

{\em Multi-particle system.}
The calculation can be easily 
extended to multi-particle systems in a grand
canonical description, i.e., at fixed temperature $T$ and chemical
potential (Fermi energy) $\mu$. Results for the number and momentum densities 
follow by replacing the probability for eigenstate
$j$ with the average occupation number, i.e.,
$e^{-E_j/T}/Z \to\ \gamma / [e^{(E_{j} - \mu)/T} + a]$,
where $a = 1$ for fermions, $-1$ for bosons, and $0$ for Boltzmann statistics,
with inert internal degrees of freedom accounted for via the degeneracy 
factor $\gamma$.

With the additional scale $\mu$, departure from classical
behavior occurs not only when $\omega_i \gton T$ 
(discrete levels matter) but also when $\mu \gg T$ (Bose/Fermi statistics is
important). The momentum anisotropy discussed here is due to the former.
Classical phase space integrals,
outlined at the end of the Introduction, give zero
anisotropy even with Bose/Fermi distribution instead of Boltzmann.

\begin{figure}
\begin{center}
\hspace*{-2mm}\includegraphics[height=40mm]{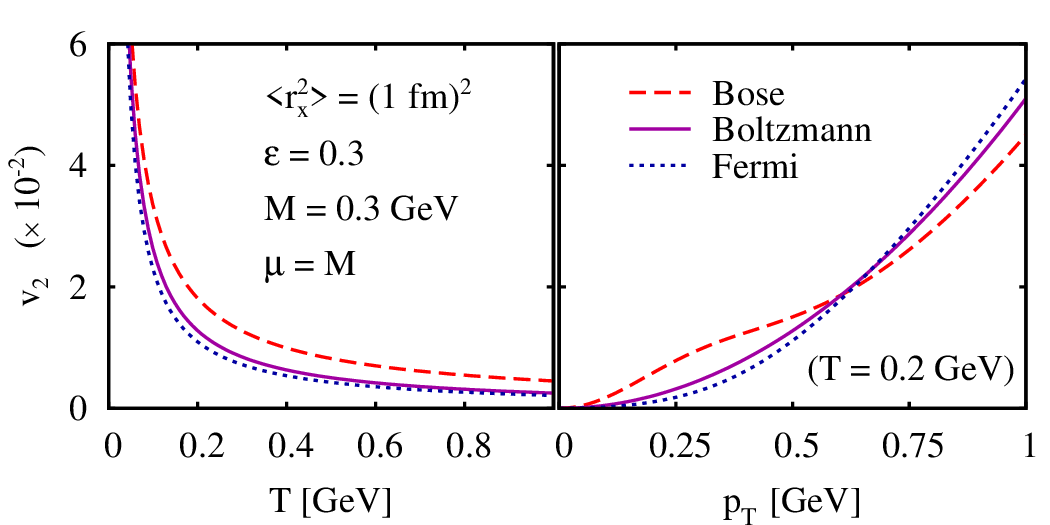}
\end{center}
\vspace*{-0.5cm}
\caption{{\em Left:} 
Average elliptic anisotropy versus temperature.
{\em Right:} Elliptic anisotropy versus transverse momentum. 
Both plots are for a multiparticle system, 
with one species of mass $M=0.3$~GeV, 
mean-squared system size $\langle r_x^2\rangle=1$~fm$^2$, spatial eccentricity
$\varepsilon=0.3$, and chemical potential $\mu = M$. 
For the right panel, $T=0.2$~GeV is used.} 
\label{fig:avv2_v2pt_multi}
\end{figure}

Figure~\ref{fig:avv2_v2pt_multi} shows the temperature and 
transverse momentum dependence of the elliptic anisotropy for multiparticle
systems at $\mu = M$, for the same $M=0.3$~GeV 
as in Fig.~\ref{fig:avv2_vs_L_eps}.
As expected, $\bar v_2$ in the left plot decreases monotonically 
with temperature, and the anisotropy is noticeably higher for Bose statistics
than for Boltzmann or Fermi statistics. This is because lower-energy states 
are more anisotropic (both spatially and in momentum space) on average.

The right plot shows the $p_T$-differential elliptic anisotropy. 
Boltzmann statistics
leads to the same analytic result (Eq.~(\ref{vn_HO})) for any $\mu$. At $\mu \approx M$,
Bose(Fermi) gives
considerable(modest) deviations in $v_2$ compared to Boltzmann.
The curves differ even at high $p_T$ because the relative weights of various energy states are different among the three types of statistics. 
Therefore, depending on statistics, 
slightly different trap parameters $\omega_i$ are needed 
in order to have the same system dimensions $\langle r_i^2\rangle^{1/2}$.
At $\mu \approx 0$ appropriate for heavy-ion collisions, deviations from Boltzmann statistics
are much smaller. Nevertheless, for all three statistics, there is considerable
``intrinsic'' elliptic anisotropy at intermediate and high $p_T$.

The same features are present for the infinite square well potential,
with even larger anisotropies at low and intermediate $p_T$ than for
the oscillator potential. At high momenta
$v_2$ saturates below unity for the square well, unlike for the
oscillator potential, because the momentum distribution has a power-law tail.

To illustrate how this intrinsic anisotropy from quantum physics might be relevant for heavy-ion collisions, Fig.~\ref{fig:rhicv2} shows the $p_T$ dependences of $v_2$ for pions and protons in the left panel and the impact parameter dependence of $\bar v_2$ for pions in the right panel, using the Au+Au collision initial geometry. The binary collision transverse profile sets the mean-squared dimensions. The left panel corresponds to Au+Au geometry at impact parameter $b=8$~fm: $\langle r_x^2\rangle^{1/2}=1.5$~fm and $\langle r_y^2\rangle^{1/2}=2.2$~fm. 
While the pion $v_2$ is significant, especially from the nonrelativistic calculation (solid curves), the proton $v_2$ is almost zero at low $p_T$. Mass ordering of $v_2$ at low $p_T$ is typical of hydrodynamics.
The quantum anisotropy shows the same qualitative 
mass ordering in the nonrelativistic limit albeit a quantitatively stronger one, with $v_2(p_T) \sim p_T^2/M^2$ at fixed $T$ and geometry. 

For massless particles (dashed curves), instead of monotonic rise with $p_T$,
one numerically finds that $v_2$ peaks near $p_T \sim T$ and gradually
decreases to a nonzero value at high $p_T$ (about 0.5\% at $b=8$ fm). 
It may be tempting to interpolate between the nonrelativistic result at low $p_T$ and the ultrarelativistic
one at high $p_T$, however, it is not clear if such a simple kinematic consideration is even qualitatively correct. 
The momentum distribution (Eq.~(\ref{f2D})), 
and hence $v_2(p_T)$, reflects the spectrum of the Hamiltonian
{\em and} its eigenstates in momentum representation --  
both of which depend on $K(\vp)$. 
A quantitative study will have to consider in the future unapproximated $K=(\vp^2 + M^2)^{1/2}$.

The pion $\bar v_2$, shown in Fig.~\ref{fig:rhicv2} right panel, may not be small compared to the measured typical $\bar v_2\approx 5\%$ in Au+Au collisions. 
In addition to $\varepsilon$, the quantum effect also depends on the size of the collision zone, i.e., $v_2$ quickly dies off towards central, small impact parameter collisions. This differs from hydrodynamics where the impact parameter dependence is nearly proportional to eccentricity. At large $b$ the collision approaches the limiting case of nucleon-nucleon interactions, where spatial anisotropy disappears in our implementation of the geometry. This explains the decreasing trend of $\bar v_2$ at large impact parameters.
The reason why the pion $\bar v_2(b)$ is about $3\times$ smaller in the massless case is that, 
as mentioned earlier, 
the massless result matches the anisotropy for a particle with mass $M \approx 2.5 T$, i.e., about $3.5 M_\pi$.

{\em Anisotropy from classical fields.}
Finally, note that a similar anisotropy effect should arise 
in any wave
dynamics. 
E.g., consider a solution to the source-free Maxwell
equations. At $t = 0$ we can write
\begin{equation}
\pmatrix{ \vE(\vr) \cr \vB(\vr)} = 
\myRe \left[\int d^3 p\, \frac{\sqrt{\omega_p}}{2\pi}\, 
\pmatrix{ \vC(\vp) \cr \vC_B(\vp)}
e^{i\vp \vr} \right]
\end{equation}
where $\omega_p \equiv |\vp|$, $\vp \cdot \vC(\vp) \equiv 0$, and
$\vC_B(\vp) = \vp \times \vC(\vp) / \omega_p$.
The total electromagnetic energy and momentum are
\begin{eqnarray}
&&\int d^3r\, \frac{1}{8\pi}[E^2(\vr) + B^2(\vr)]
= \int d^3p\, \omega_p\, |\vC(\vp)|^2 \,,
\qquad
\\
&&
\int d^3 r\, \frac{1}{4\pi} \vE(\vr) \times \vB(\vr) 
  = \int d^3p\, \vp\,|\vC(\vp)|^2 \,,
\end{eqnarray}
so we can interpret
$f(\vp) = |\vC(\vp)|^2$ 
as the momentum distribution of (on-shell) photons.
If the energy density is spatially anisotropic,
in general
the momentum distribution will be anisotropic as well.
It would be interesting to check to what extent differential elliptic anisotropy
from the classical Yang-Mills ``color glass'' approach~\cite{NaraKrasnitz_v2}
can be attributed to the Fourier transform of the initial shape of the
overlap region.

{\em Discussions and Conclusions.} 
We conclude that quantum mechanics can be relevant for the momentum anisotropy in hadron and nuclear collisions. 
Our estimates
indicate that the effect is significant and may not be neglected. 
Our results
 are necessarily qualitative: 
{\em i)} we used simple harmonic oscillator and nonrelativistic quantum 
mechanics but also corroborate many of our findings with calculations for massless particles; 
{\em ii)} we assumed a thermalized system at temperature $T\sim 0.2$~GeV; 
{\em iii)}
we used a single statistical ensemble--what if we only have {\em local} thermalization\cite{global_therm}, i.e., many small thermal systems distributed spatially? 
{\em iv)} We used $M \sim 0.3$~GeV with constituent quarks in mind--will the resulting hadron anisotropy be actually larger in a coalescence picture~\cite{coal}? 
{\em v)} We estimated pion and proton $v_2$ with a bag of hadrons in mind at initial Au+Au encounter--how would subsequent dynamical evolution such as hydrodynamic expansion affect the anisotropy? 
It is not clear, even qualitatively, 
what fraction of the initial quantum anisotropy remains after hydrodynamic 
expansion -- which is presumably needed, in addition, to produce the measured
final-state anisotropy --  or how strong quantum effects would be
if we imposed quantum uncertainty right 
at the late breakup stage of hydrodynamic 
evolution.
Many of the open questions concern how to properly include
quantum mechanics in relativistic, dissipative hydrodynamics.
But the important message of our work is that quantum physics must be present 
and needs rigorous consideration, 
especially for small systems such as proton-proton and 
proton-nucleus collisions.

Strong elliptic anisotropy was also observed in cold atom systems released from anisotropic trap~\cite{OHara}. For those systems, the quantum mechanics effect is minuscule and, therefore, the elliptic anisotropy is due to interactions during expansion. E.g., the experiment by O'Hara {\em et al.}~\cite{OHara} confined fermionic lithium atoms at $T\sim 1$~$\mu$K, Fermi temperature $T_F \sim 10T$, in a trap of dimensions $r_x\sim 20$~$\mu$m and $r_y\sim 100$~$\mu$m, for which we estimate using Eq.~(\ref{HOv2_approx}) a quantum anisotropy $\bar v_2\sim10^{-5}$. It is small because the cold atom system's intrinsic momentum quantum $\hbar/r_x\sim10^{-8}$~MeV$/c$
is negligible compared to the typical momentum $\sqrt{k_B T M} \sim 10^{-6}$~MeV$/c$. The intrisic energy quantum $\hbar^2/M r_x^2 \sim 10^{-20}$~MeV is also much smaller than the temperature $k_B T \sim 10^{-16}$~MeV. In this sense these cold atoms are much ``hotter'' than the hot quark gluon plasma, where the temperature is comparable to the intrinsic momentum and energy scales. The cold atom system is thus classical in its expansion dynamics (despite being in the 
$S$-wave scattering limit),
while the quark gluon plasma may exhibit intrinsic quantum features. It would be interesting to do an experiment with cold lithium atoms in a 
$\sim 100$ times smaller trap or at nano-Kelvin temperatures
(conditions closer to the few-body limit~\cite{Serwane}),
or a trapped cold electron system.

{\em Acknowledgments.} This work was supported in part by the U.S. Department of 
Energy, Office of Science, under 
Awards No. DE-SC0004035 and DE-SC0016524 (DM), DE-SC0010545 (CHG), and DE-SC0012910 (FW).



\begin{thebibliography}{99}
\bibitem{note_z} This can correspond to uniform
distribution in the $z$ direction (for single-particle system),
or instead, tight confinement in the $z$ direction.
\bibitem{note_unit} We use high-energy physics units 
$\hbar = c = k_B = 1$.
\bibitem{note_rep} This is because, for harmonic oscillator, 
up to constant scaling factors
the Schr\"odinger equation reads the same in both coordinate
and momentum representation.
\bibitem{note_gaus}
$\partial_\phi f = 0$ implies
$p_y\, G'(p_x) \equiv p_x\, G'(p_y)$ with $G(p) = \ln g(p)$.
Substituting $p_y = p_x + \delta$ and Taylor expanding in $\delta$
gives the differential equation $p\, G''(p) = G'(p)$ 
with solution $G'(p) \propto p$, so 
$g(p)$ is Gaussian.
\bibitem{global_therm}
Global equilibrium is inconsistent with causality
in early stages of the collision, and also with the nonuniform
fluid velocity field of hydrodynamic solutions in later stages.
\end{thebibliography}
\end{document}